\documentclass[journal,twoside]{IEEEtran}
\usepackage[caption=false,font=footnotesize]{subfig}
\usepackage[cmex10]{amsmath}

\usepackage{cite}

\ifCLASSINFOpdf
\usepackage[pdftex]{graphicx}
\else
\usepackage[dvips]{graphicx}
\fi
\usepackage{multirow}

\usepackage[cmex10]{amsmath}
\usepackage{amsfonts}
\usepackage[export]{adjustbox}
\usepackage{xcolor}
\usepackage{algorithmic}

\usepackage{array}
\usepackage{multicol}
\usepackage{colortbl}
\usepackage{footnote}
\usepackage{xfrac}

\ifCLASSOPTIONcompsoc
 \usepackage[caption=false,font=normalsize,labelfont=sf,textfont=sf]{subfig}
\else
\usepackage[caption=false,font=footnotesize]{subfig}
\fi
\usepackage{fixltx2e}
\usepackage{dblfloatfix}
\usepackage{url}
\usepackage[colorlinks=true, citecolor=green, linkcolor=red, urlcolor=blue]{hyperref}
\usepackage[hyperpageref]{backref}
\usepackage{stackengine}

\newcommand{\Conv}{\mathop{\scalebox{1.5}{\raisebox{-0.2ex}{$\ast$}}}}

\newcolumntype{N}{>{\centering\arraybackslash}m{2.5cm}}
\newcolumntype{M}{>{\centering\arraybackslash}m{2.4cm}}
\newcolumntype{A}{>{\centering\arraybackslash}m{1.9in}}
\newcolumntype{B}{>{\centering\arraybackslash}m{4.5in}}
\newcolumntype{C}{>{\centering\arraybackslash}m{0.57in}}

\makeatletter

\newcommand{\Rmnum}[1]{\expandafter\@slowromancap\romannumeral #1@}

\makeatother
\DeclareMathOperator*{\argmax}{argmax}

\newcommand\Alpha{\mathrm{A}}
\newcommand\Beta{\mathrm{B}}
\DeclareMathOperator{\Tr}{Tr}
\def\delequal{\mathrel{\ensurestackMath{\stackon[1pt]{=}{\scriptstyle\Delta}}}}
\def\dotequal{\mathrel{\ensurestackMath{\stackon[1pt]{=}{\scriptstyle\cdot}}}}

\begin{document}
\bstctlcite{IEEEexample:BSTcontrol}
%
\title{Lung Cancer Lesion Detection in Histopathology Images Using Graph-Based Sparse PCA Network}
%
%
%
\author{Sundaresh~Ram,~\IEEEmembership{Member,~IEEE,} Wenfei~Tang, Alexander~J.~Bell, Cara~Spencer, Alexander~Buschhaus, Charles~R.~Hatt,~\IEEEmembership{Member,~IEEE}, Marina~Pasca~diMagliano, Jeffrey~J.~Rodr\'iguez,~\IEEEmembership{Senior~Member,~IEEE}, Stefanie~Galban, and Craig~J.~Galban
\thanks{Manuscript received Month XX, 20xx; revised Month xx 20xx; accepted Month 20xx. This work was supported in part by the National Heart, Lung, and Blood Institute of the National Institute of Health under Grant R01HL139690 and in part by grant support awarded to Dr. Stefanie Galban by the Rogel Cancer Center at the University of Michigan. \textit{(corresponding author: Sundaresh Ram)}}
\thanks{Sundaresh Ram, Alexander J. Bell, and Craig J. Galban are with the Department of Radiology, and the Department of Biomedical Engineering, University of Michigan, Ann Arbor, MI, 48109 USA (e-mail: sundarer@umich.edu, cgalban@umich.edu).}
\thanks{Wenfei Tang is with the Department of Computer Science and Engineering, University of Michigan, Ann Arbor, MI, 48109 USA.}
\thanks{Cara Spencer is with the Department of Computational Medicine and Bioinformatics, University of Michigan, Ann Arbor, MI, 48109 USA.}
\thanks{Charles R. Hatt is with Imbio LLC, Minneapolis, MN, 55405 USA, and with the Department of Radiology, University of Michigan, Ann Arbor, MI, 48109 USA (e-mail: chatti@umich.edu).}
\thanks{Marina Pasca diMagliano is with the Department of Surgery and Department of Cell and Developmental Biology, University of Michigan, Ann Arbor, MI, 48109 USA.}
\thanks{Jeffrey J. Rodr\'iguez is with the Department of Electrical and Computer Engineering, and Department of Biomedical Engineering, The University of Arizona, Tucson, AZ 85721, USA (e-mail: jjrodrig@arizona.edu).}
\thanks{Alexander Buschhaus, Stefanie Galban are with the Department of Radiology, University of Michigan, Ann Arbor, MI, 48109 USA.}}
\maketitle


\begin{abstract}
 Early detection of lung cancer is critical for improvement of patient survival. To address the clinical need for efficacious treatments, genetically engineered mouse models (GEMM) have become integral in identifying and evaluating the molecular underpinnings of this complex disease that may be exploited as therapeutic targets. Assessment of GEMM tumor burden on histopathological sections performed by manual inspection is both time consuming and prone to subjective bias. Therefore, an interplay of needs and challenges exists for computer-aided diagnostic tools, for accurate and efficient analysis of these histopathology images. In this paper, we propose a simple machine learning approach called the graph-based sparse principal component analysis (GS-PCA) network, for automated detection of cancerous lesions on histological lung slides stained by hematoxylin and eosin (H\&E). Our method comprises four steps: 1) cascaded graph-based sparse PCA, 2) PCA binary hashing, 3) block-wise histograms, and 4) support vector machine (SVM) classification. In our proposed architecture, graph-based sparse PCA is employed to learn the filter banks of the multiple stages of a convolutional network. This is followed by PCA hashing and block histograms for indexing and pooling. The meaningful features extracted from this GS-PCA are then fed to an SVM classifier. We evaluate the performance of the proposed algorithm on H\&E slides obtained from an inducible \textit{K-ras}$^{\textit{G12D}}$ lung cancer mouse model using precision/recall rates, $F_{\beta}$-score, Tanimoto coefficient, and area under the curve (AUC) of the receiver operator characteristic (ROC) and show that our algorithm is efficient and provides improved detection accuracy compared to existing algorithms.  
\end{abstract}


%
\IEEEpeerreviewmaketitle

\section{Introduction}
\label{sec:intro}

\begin{figure}[!t]
\includegraphics[width= 3.5in, height=1.826in]{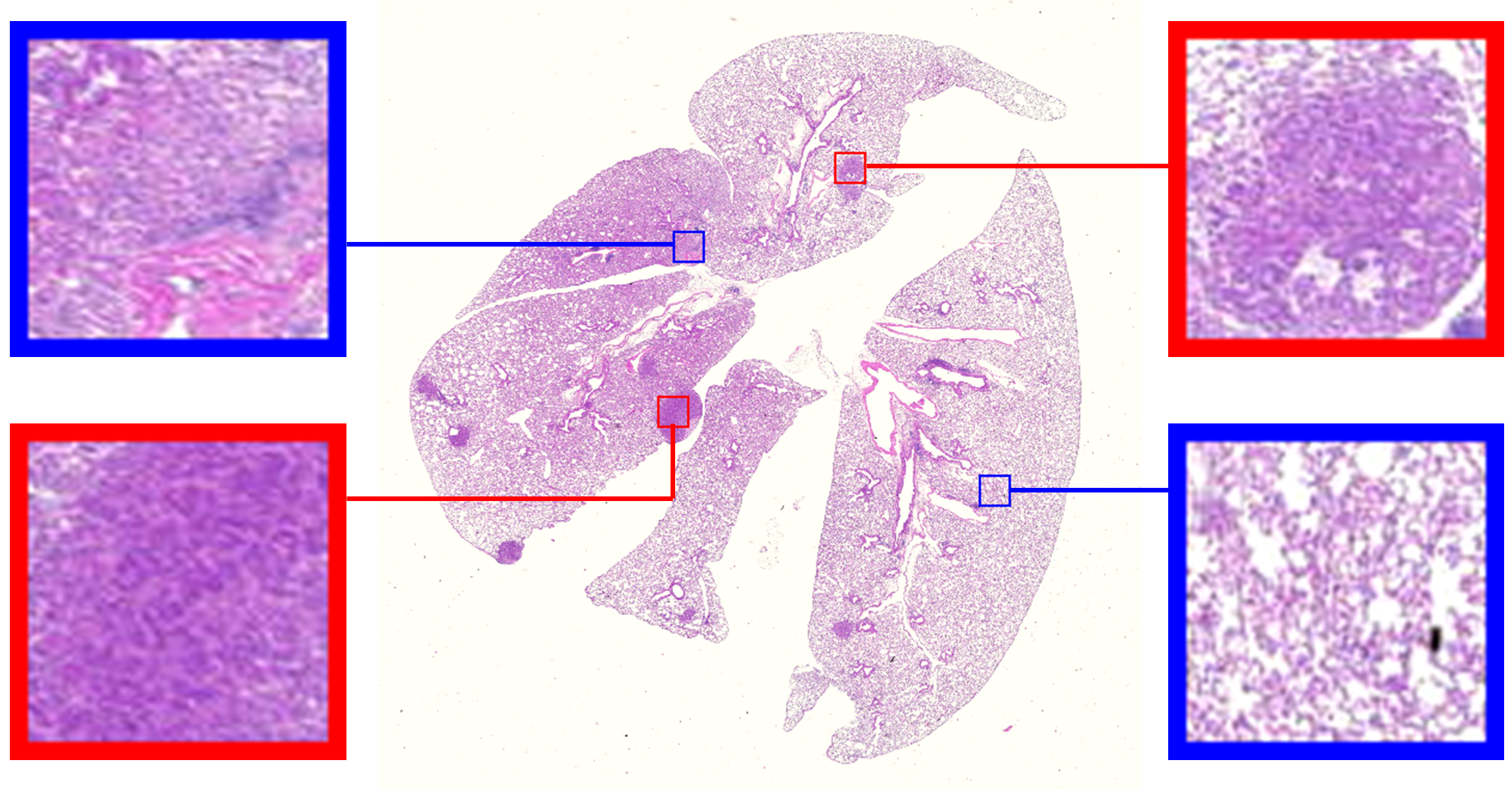}
\caption{An example whole-slide histopathological image from our dataset consisting of many tumor lesions. The high-resolution inset images show the visual features that characterize the tumor (red frame) and normal (blue frame) regions.}
\label{fig1}
\vspace{-4mm}
\end{figure}

\IEEEPARstart {L}{ung} cancer is the leading cause of cancer-related deaths worldwide, with an estimated 1.6 million deaths each year \cite{cruz11lung}. Development of novel therapies to battle lung cancer has been greatly aided by the emergence of genetically engineered mouse models (GEMMs) of lung cancer, such as the \textit{K-ras}$^{\textit{G12D}}$; \textit{p53}$^{\textit{Frt}}$ non–small-cell lung carcinoma (NSCLC) model, where the compound effect of conditional mutations in the \textit{K-ras} oncogene and the \textit{p53} tumor suppressor gene leads to development of adenocarcinomas in the mouse lung \cite{walrath10genetically, barck15quantification}. Since GEMMs recapitulate certain aspects of the human disease associated with the stroma, vascularity, and immune infiltrate better than other models, it is important to be able to detect, identify and localize the lung tumor lesions seen on the histopathological sections as shown in Fig.~\ref{fig1}. 

Manual assessment of tumor burden (the amount of tumor cells/mass present in a subject's body) on histopathological mouse lung sections is difficult, time consuming, and a labor-intensive process. This is due to various reasons such as fluctuating intensities \cite{ram13symmetry}, color change and morphological variations within structures of the cancer lesions in these images \cite{lin19fast}, tumor heterogeneity \cite{junttila13influence} (see Fig.~\ref{fig1}), low signal-to-noise-ratio \cite{ram10seg,ram16size}, variations in illumination \cite{ram18three}, microscopy imaging limitations \cite{ram12size, ram2017sparse, ram18classify, ram20combined}, and the large number of images and the number of lesions per image an expert has to demarcate. Moreover, the task of manual detection of cancer lesions on H\&E slides can be subjective, leading to inter-observer variability. Therefore, there is a pressing need for computer-aided diagnostic tools for accurate and efficient quantitative analysis of histopathology images \cite{gurcan09histo,veta14breast,xing16robust, ram21detect}.

Tumor detection and classification tools within the commonly available microscopy software are based on feature extraction techniques such as size, shape, and morphological features \cite{gurcan09histo, basavanhally13multi, gorelick13pros, veta14breast, xing16robust, ram16size, tizhoosh18represent}, texture features including local binary pattern (LBP) \cite{reis17auto, wan17integrated, simon18multi}, local Fourier transform \cite{kong11parti}, co-occurrence matrix and fractal texture features \cite{alinsaif20part}, and energy minimization and optimization-based techniques \cite{tosun11graph, ozdemir13hybrid, bejnordi16automated, javed20multi}. These techniques suffer considerably due to over-generalization and therefore need extensive customization for the dataset at hand, limiting their use to very simple images obtained/collected in a carefully constrained environment \cite{ram16size}. Tumor detection and grading using size, shape and other morphological features does not work well when the cell population exhibits a variety of sizes and shapes, or when the signal-to-noise (SNR) ratio is poor \cite{shi17histo}. Energy minimization and optimization techniques minimize the internal energy within tumor areas for their accurate detection, but may lead to false detections for highly textured and heterogeneous tumor lesions. To overcome these limitations, existing software tools allow user-friendly interfaces to correct the results obtained. This, however, results in losing the benefits of automation such as speed and reproducibility.

There has been much interest in developing algorithmic methods that adapt naturally to the dataset and perform feature discovery. One such popular class of learning or feature discovery methods includes those based on sparse representation-based classification (SRC) \cite{wright09robust}. There have been many SRC methods that have been successfully applied to a variety of histopathological image classification problems \cite{srinivas14simul, vu16histo, sarkar18sdl, li20anal}. These methods are based on finding linear representations in the data. However, linear representations are almost always inadequate for representing non-linear structures of the data which arise in many practical applications. A recent class of learning-based methods involve the design of deep neural networks that can be trained to learn relevant features by themselves. There have been plenty of deep learning methods that have been developed for histopathological image classification \cite{hou16patch, xu17large, tellez18whole, lin19fast, xing19pixel, campanella19clinical, wei19patho, valkonen20cyto}. The success of deep learning, however, has been fueled by the availability of generous and clean training data. When the training data is limited and/or noisy, as is often the case in medical imaging, these methods tend to show a performance degradation \cite{goodfellow16deep}. Another class of learning-based approaches involve orthogonal transformation of the data such as principal component analysis (PCA) transform to extract relevant features for image classification \cite{chan15pcanet, bruna13invariant, shi17histo, dutta20sparse}. These learning-based approaches using orthogonal transformation explore the data distribution to preserve global structures in the data.   

In this paper, we present a simple machine learning approach called the graph-based sparse principal component analysis (GS-PCA) network, which combines the local and global structures of all the data and is implemented in a deep learning framework to learn an explicit nonlinear mapping of the data for accurate detection and classification. We use the most basic and easy operations to emulate the processing stages in a typical (convolutional) neural network: First, graph-based sparse PCA filters are used as the data-adapting convolutional filter bank at each stage of the network. Next, we perform a simple binary quantization (hashing) that serves as the nonlinear stage, followed by block-wise histograms of the binary codes as the feature pooling stage to obtain the final output features of the network. Finally, we train a support vector machine (SVM) classifier on the output features of the network to obtain the final classification instead of the regular softmax classifier, as the softmax classifier known to overfit \cite{chan15pcanet}. For ease of reference, we call this data-processing network a \emph{Graph-based Sparse PCA Network} (GS-PCANet). The key contributions of this paper are as follows:
\begin{itemize}
    \item \textbf{Feature Extraction Using Graph-Based Sparse PCA:} Unlike other histopathology image classification methods, in this work we propose a baseline neural network method called GS-PCANet, which is different from prior methods \cite{bruna13invariant, chan15pcanet, shi17histo, dutta20sparse} in two aspects. 1) We include an additional sparsity promoting term in the PCA transformation so as to select more interpretable features from the images. 2) We include a graph regularization term in the objective function so as to preserve the local structures for each data point between the different classes.  
    \item \textbf{Computationally Efficient Approach:} Our proposed GS-PCANet is computationally efficient in comparison to other deep learning methods in two aspects. 1) We show that a simple two-stage network is good enough to extract all the relevant features for classifying the tumor versus healthy lung regions. 2) We do not need to learn the filter weights at each stage of the network. 
\end{itemize}
We evaluate the proposed method and seven state-of-the-art algorithms developed for histopathology image classification on a dataset of 67 images provided by the Stefanie Galban Lab, at the University of Michigan. The dataset consists of microscopy images of murine H\&E stained lung sections and are divided into two categories: images of non-tumor-bearing control mice and images of mice with visible tumor.
%
\begin{figure*}[!t]
\includegraphics[width= 7in, height=2.15in]{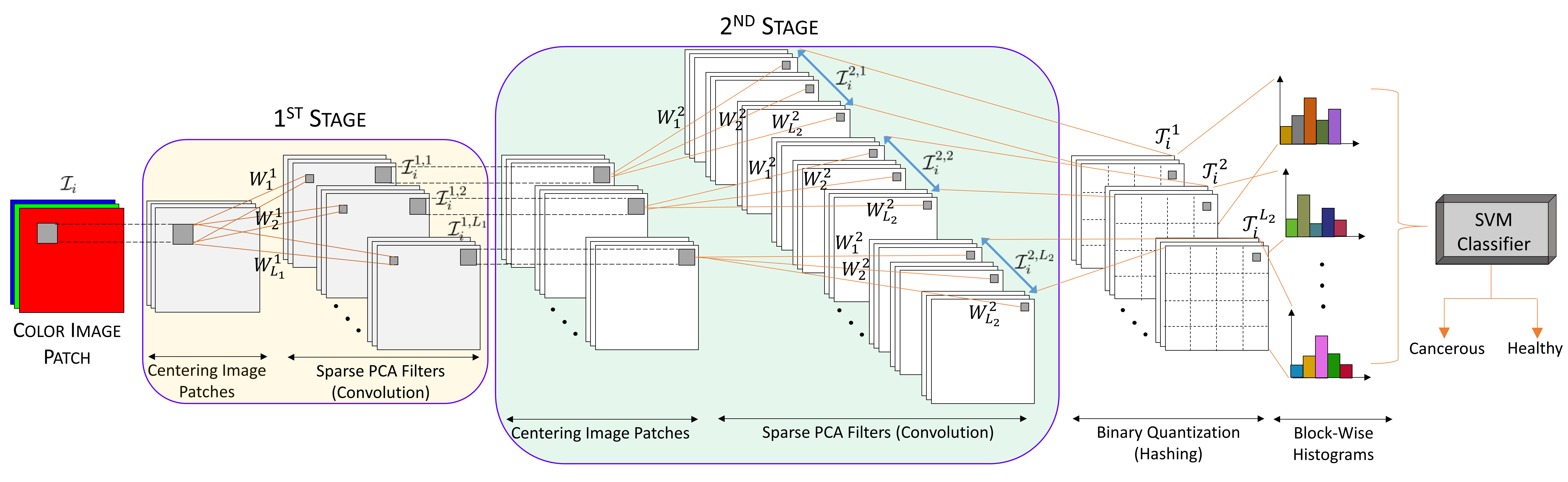}
\caption{An outline of the proposed (two-stage) GS-PCANet.}
\label{fig2}
\vspace{-5mm}
\end{figure*}

\section{Principal Component Analysis}
\label{sec:pca}

Let $\boldsymbol{\text{X}}$ denote an $n \times p$ matrix of $n$ rows and $p$ columns of rank $q \leq \text{min}\left(n, p\right)$, where $n$ is the number of data samples, and $p$ is the number of features/variables. Let $\boldsymbol{\text{x}}_{i}^{j}$ denote the element of $\boldsymbol{\text{X}}$ at row $i$ and column $j$. Assume each column has zero mean. Let $\Sigma$ denote the covariance matrix of $\boldsymbol{\text{x}}_{i}$, where $\Sigma$ is a positive definite matrix of size $p \times p$, which can be decomposed as
\begin{equation}
\label{eq:1}
    \Sigma = \sum_{i = 1}^{q} \sigma_{i}\boldsymbol{\text{v}}_{i}\boldsymbol{\text{v}}_{i}^{\top}
\end{equation}
where $\sigma_{i}$ is the $i^{\text{th}}$ largest eigenvalue of $\Sigma$ and $\boldsymbol{\text{v}}_{i} = \left[v_{i1}, \hdots, v_{ip}\right]^{\top}$ is its associated eigenvector. PCA reduces the dimensionality of the data from $p$ to $q$ by replacing the original features/variables with $q$ linear combinations of the form $\boldsymbol{\text{X}}\boldsymbol{\text{v}}_{k},\, k = 1, \hdots, q$ known as the principal components (PCs), which are obtained by maximizing their variance:
\begin{equation*}
    \boldsymbol{\text{v}}_{k} = \argmax_{\boldsymbol{\text{v}}} \big\{\text{Var}\left(\boldsymbol{\text{Xv}}\right)\big\} \quad \text{subject to} \quad \boldsymbol{\text{v}}_{k}^{\top} \boldsymbol{\text{v}}_{k} = 1
\end{equation*}
and
\begin{equation*}
    \boldsymbol{\text{v}}_{j}^{\top} \boldsymbol{\text{v}}_{k} = 0 \quad \text{for} \quad j < k
\end{equation*}
where $\boldsymbol{\text{v}}_{k}$ is the $k^{\text{th}}$ principal loading vector and the projection of the data $\boldsymbol{\text{X}}\boldsymbol{\text{v}}_{k}$ is the $k^{\text{th}}$ principal component and the operator $\text{Var}\left(\cdot\right)$ denotes the (estimated) variance of a random variable.

Generally, PCA is computed using singular value decomposition (SVD) of $\boldsymbol{\text{X}}$ as
\begin{equation}
\label{eq:2}
    \boldsymbol{\text{X}} = \boldsymbol{\text{USV}}^{\top}
\end{equation}
where the columns of $\boldsymbol{\text{Z}} \delequal \boldsymbol{\text{US}}$ are the PCs, and the columns of $\boldsymbol{\text{V}}$ are the corresponding principal loading vectors (also known as basis vectors) \cite{malladi20image}. The matrix $\boldsymbol{\text{S}}$ is a $q \times q$ diagonal matrix of ordered singular values $s_{1} \geq s_{2} \geq \hdots \geq s_{q} > 0$ and the columns of $\boldsymbol{\text{U}}$ and $\boldsymbol{\text{V}}$ are orthonormal such that $\boldsymbol{\text{U}}^{\top}\boldsymbol{\text{U}} = \boldsymbol{\text{V}}^{\top}\boldsymbol{\text{V}} = \boldsymbol{\text{I}}_{q}$. If $\boldsymbol{\text{X}}$ is low rank, it is possible to significantly reduce its dimensionality by using the $q$ most significant basis vectors. The projection of the data $\boldsymbol{\text{X}}$ upon the first $q$ basis vectors gives the PCs.

An alternative formulation for PCA can be derived on the projection framework \cite{chan15pcanet}, where the PC loading matrix $\boldsymbol{\text{V}}$ also known as the PCA basis (defined as the matrix containing the principal loading vectors) can be estimated by solving the following least squares optimization problem:
\begin{equation}
\label{eq:3}
    \min_{\boldsymbol{\text{A}}} \big\{\parallel\boldsymbol{\text{X}} - \boldsymbol{\text{X}}\boldsymbol{\Alpha\Alpha}\!^{\top}\parallel_{F}^{2}\big\} \quad \text{subject to} \quad \boldsymbol{\Alpha}\!^{\top}\!\boldsymbol{\Alpha} = \boldsymbol{\mathbb{I}}_{q} 
\end{equation}
where $\parallel \cdot \parallel_{F}$ is the Frobenius norm, $\boldsymbol{\Alpha} \in \mathbb{R}^{p \times q}$ is a matrix whose columns form an orthonormal basis $\left\{\boldsymbol{\alpha}_{1}, \boldsymbol{\alpha}_{2}, \hdots, \boldsymbol{\alpha}_{q}\right\}$, and $\mathbb{I}_{q}$ is an identity matrix of size $q \times q$. The columns of $\boldsymbol{\Alpha}$ that minimize (\ref{eq:3}) are referred to as the PCA basis $\boldsymbol{\text{V}}$. The minimization is solved by formulating it as a least absolute shrinkage and selection operator (LASSO) problem \cite{zou06sparse}. Each principal component is derived from a linear combination of all $p$ features, consequently making $\boldsymbol{\alpha}$ non-sparse. We use this alternative formulation for PCA feature extraction in this work. 
%

\section{Proposed Method}
\label{sec:mthd}

Based on the PCA methodology, we propose a simple and efficient machine learning method for histopathology image classification. First, we obtain graph-based sparse PCA filters from the training images as the data adaptive convolutional filter bank for the various stages of a convolutional neural network. Then we perform a simple binary quantization (hashing), which serves as a nonlinear stage. Next, we use block-wise histograms of the binary codes obtained from the quantization process to get the output features of the network. Finally, we train a SVM classifier using the output features to obtain the final classification. The proposed GS-PCANet model is shown in Fig.~\ref{fig2}, illustrating each of the above steps involved in our algorithm.

\subsection{Graph-Based Sparse PCA}
\label{sec:mthda}

From the analysis of PCA in Section~\ref{sec:pca}, we can obtain a sparse PCA basis by including a regularization term in (\ref{eq:3}). Inclusion of a sparsity penalty reduces the number of features involved in each linear combination for obtaining the PCs. One way to extend (\ref{eq:3}) to obtain sparse basis vectors is by imposing $\ell_{1}$-norm and $\ell_{2}$-norm penalty constraints upon the regression coefficients (basis vectors) \cite{zou06sparse}:
\begin{equation}
\label{eq:4}
    \min_{\boldsymbol{\text{A},\, \text{B}}} \Bigg\{\parallel\!\boldsymbol{\text{X}} - \boldsymbol{\text{X}}\boldsymbol{\Beta\Alpha}\!^{\top}\!\parallel_{F}^{2} + \lambda \sum_{j = 1}^{q}\parallel\!\boldsymbol{\beta}_{j}\!\parallel_{2}^{2} + \sum_{j = 1}^{q} \lambda_{1,j}\parallel\!\boldsymbol{\beta}_{j}\!\parallel_{1} \Bigg\} 
\end{equation}
\begin{equation*}
     \text{subject to} \quad \boldsymbol{\Alpha}\!^{\top}\!\boldsymbol{\Alpha} = \boldsymbol{\mathbb{I}}_{q}
\end{equation*}
where the same $\lambda$ (the regularization parameter of the $\ell_{2}$-norm) is used for all $q$ components, different $\lambda_{1,j}\text{'s}$ (the regularization parameters of the $\ell_{1}$-norm) are allowed for penalizing the loadings of different PCs. The $\boldsymbol{\Beta} \in \mathbb{R}^{p \times q}$ corresponds to the required sparse basis $\left\{\boldsymbol{\beta}_{1}, \boldsymbol{\beta}_{2}, \hdots, \boldsymbol{\beta}_{q}\right\}$. The $\ell_{1}$-norm and $\ell_{2}$-norm regularization terms penalize the number of non-zero coefficients in $\boldsymbol{\beta}$, whereas the loss term simultaneously minimizes the reconstruction error $\parallel\!\boldsymbol{\text{X}} - \boldsymbol{\text{X}}\boldsymbol{\Beta\Alpha}\!^{\top}\!\parallel_{F}^{2}$. If $\lambda$ and the $\lambda_{1,j}\text{'s}$ are zero, the problem reduces to finding the ordinary PCA basis vectors, equivalent to (\ref{eq:3}). When $\lambda,\,\lambda_{1,j}\text{'s} > 0$ some coefficients of $\boldsymbol{\beta}_{j}$ are forced to zero, resulting in sparsity.

The sparse PCA defined in (\ref{eq:4}) preserves the global structures in the data. In addition to preserving the global structures, we are interested in preserving the local structures, i.e., $k$ nearest neighbor ($k$NN) preservation of each data point $\boldsymbol{\text{x}}_{i}$, as they help in identifying local features in the data. We define $\mathcal{M} = \big\{\boldsymbol{\text{X, E}}\big\}$ to be a constructed weighted graph. The vertices of $\mathcal{M}$ correspond to the data points $\left\{\boldsymbol{\text{x}}_{1}, \boldsymbol{\text{x}}_{2}, \hdots, \boldsymbol{\text{x}}_{n}\right\}$. The weight matrix $\boldsymbol{\text{E}} = \left(e_{lm}\right)_{l,m = 1}^{n} \in \mathbb{R}^{n \times n}$ is defined as
\begin{eqnarray}
\label{eq:5}
    e_{lm} = \left\{\begin{array}{l l}
    1, & \text{if}\quad l, m \in N_{k}(i)~\forall~ i \\
    0, & \text{otherwise} \\
    \end{array} \right.
\end{eqnarray}
where the set $N_{k}(i)$ contains the $k$ nearest neighbors to the node $i$ in the graph. Furthermore, the $\ell_{2}$-norm $\parallel\!\boldsymbol{\text{x}}_{l} - \boldsymbol{\text{x}}_{m}\!\parallel^{2}_{2}$ is applied to measure the dissimilarity of two data points, and the weight matrix $\boldsymbol{\text{E}}$ is used to restrict the similarity between two data points. Thus, with the weight matrix $\boldsymbol{\text{E}}$, we can formulate a graph regularization term as
\begin{align}
\label{eq:6}
     \frac{1}{2} \sum\limits_{l,m = 1}^{n}\!e_{lm}& \parallel\!\boldsymbol{\text{x}}_{l} - \!\boldsymbol{\text{x}}_{m}\!\parallel_{2}^{2} ~= \sum\limits_{k = 1}^{n} \boldsymbol{\text{x}}_{l}^{\top} e_{mm}\boldsymbol{\text{x}}_{l} -\! \sum\limits_{l,m = 1}^{n} \boldsymbol{\text{x}}_{l}^{\top} e_{lm} \boldsymbol{\text{x}}_{l} \nonumber \\ & = \Tr(\boldsymbol{\text{XCX}}^{\top}) - \Tr(\boldsymbol{\text{XEX}}^{\top}) = \Tr(\boldsymbol{\text{XLX}}^{\top}) 
\end{align} 
where $\boldsymbol{\text{C}}$ is a diagonal matrix with $\boldsymbol{\text{C}}_{ll} = \sum_{m = 1}^{n} e_{lm}$, $\boldsymbol{\text{L}}$ is the graph Laplacian matrix computed as $\boldsymbol{\text{L}} = \boldsymbol{\text{C}} - \boldsymbol{\text{E}}$ and $\Tr$ is the trace of a matrix. Minimizing the graph regularization term in (\ref{eq:6}) ensures that the local structures between the data points are preserved. Combining the sparse PCA from (\ref{eq:4}) and the graph regularization from (\ref{eq:6}), we propose a graph-based sparse PCA model,
\begin{align}
\label{eq:7}
    \min_{\boldsymbol{\text{A},\, \text{B}}} \Bigg\{\parallel\!\boldsymbol{\text{X}}  - \boldsymbol{\text{X}}\boldsymbol{\Beta\Alpha}\!^{\top}\!& \parallel_{F}^{2} ~ + ~ \lambda \sum_{j = 1}^{q}\parallel\!\boldsymbol{\beta}_{j}\!\parallel_{2}^{2} \\
    & + \sum_{j = 1}^{q} \lambda_{1,j}\parallel\!\boldsymbol{\beta}_{j}\!\parallel_{1} ~ + ~ \rho \Tr(\boldsymbol{\text{XLX}}^{\top}) \Bigg\} \nonumber
\end{align}
\begin{equation*}
     \text{subject to} \quad \boldsymbol{\Alpha}\!^{\top}\!\boldsymbol{\Alpha} = \boldsymbol{\mathbb{I}}_{q}
\end{equation*}
where $\rho$ is a graph regularization parameter. To solve (\ref{eq:7}), we perform the following steps: first solve an ordinary PCA problem to fix $\boldsymbol{\text{A}}$, then formulate an elastic net with the fixed $\boldsymbol{\text{A}}$ and solve for $\boldsymbol{\text{B}}$, then perform SVD to update $\boldsymbol{\text{A}}$, and repeat these steps until convergence, finally obtaining the solution as $\boldsymbol{\text{V}=\text{B}/\parallel \text{B}\parallel}$. 

\subsection{Architecture of GS-PCA Network}
\label{sec:mthdb}

Suppose there are $N$ training images $\left\{I_{i}\right\}_{i = 1}^{N}$ of size $u \times v$, and assume that PCA filter size is $t_{1} \times t_{2}$ (formed by reshaping a basis vector of length $t_1\times t_2$) at all stages of the network. The sparse PCA filters are learned from these training images. We describe each component of the network in detail below (see Fig.~\ref{fig2}).

\subsubsection{First Stage (GS-PCA)} 
For each training image $I_{i}$, around each pixel we take an image patch of size $t_{1} \times t_{2}$ and denote all the overlapping image patches in the $i^{\text{th}}$ image as $\boldsymbol{\text{X}}_{i} = \left[\boldsymbol{\text{x}}_{i,1}, \boldsymbol{\text{x}}_{i,2}, \hdots, \boldsymbol{\text{x}}_{i,\Tilde{u}\Tilde{v}}\right]$, where $\boldsymbol{\text{x}}_{i,j}$ denotes the $j^{\text{th}}$ vectorized image patch in $I_{i}$, $\Tilde{u} = u - \left(t_{1} - 1\right)$, $\Tilde{v} = v - \left(t_{2} - 1\right)$. We then subtract the image patch mean from each of the image patches and obtain the centralized matrix $\boldsymbol{\Bar{\text{X}}}_{i}$ of $\boldsymbol{\text{X}}_{i}$ as $\boldsymbol{\Bar{\text{X}}}_{i} = \left[\boldsymbol{\Bar{\text{x}}}_{i,1}, \boldsymbol{\Bar{\text{x}}}_{i,2}, \hdots, \boldsymbol{\Bar{\text{x}}}_{i,\Tilde{u}\Tilde{v}}\right]$, where $\boldsymbol{\Bar{\text{x}}}_{i,j} = \boldsymbol{\text{x}}_{i,j} - \mu_{i}$ and $\mu_{i} = \mathbb{E}\,\left[\boldsymbol{\text{x}}_{i}\right] \approx (\frac{1}{n})\sum\limits_{j = 1}^{n} \boldsymbol{\text{x}}_{i,j}$. By constructing a similar centralized matrix for each training image $I_{i}$, we obtain
\begin{equation}
\label{eq:8}
    \boldsymbol{\text{X}} = \left[\boldsymbol{\Bar{\text{X}}}_{1}, \boldsymbol{\Bar{\text{X}}}_{2}, \hdots, \boldsymbol{\Bar{\text{X}}}_{N}\right] \in \mathbb{R}^{t_{1}t_{2} \times N\Tilde{u}\Tilde{v}}.
\end{equation}
Assuming that we have $L_{i}$ PCA filters in stage $i$, sparse PCA minimizes the reconstruction error within a family of orthonormal filters using (\ref{eq:7}), where $\boldsymbol{\mathbb{I}}_{q}$ is an identity matrix of size $L_{1} \times L_{1}$. The solution to the minimization problem in (\ref{eq:7}) are the $L_{1}$ principal eigenvectors of $\boldsymbol{\text{XX}}^{\top}$ \cite{chan15pcanet}. The PCA filters can therefore be expressed as
\begin{equation}
\label{eq:9}
    \boldsymbol{W}_{l_{1}}^{1} \dotequal \text{Mat}_{t_{1},t_{2}}\left[\mathcal{Q}_{l_{1}}\left\{\boldsymbol{\text{XX}}^{\top}\right\}\right] \in \mathbb{R}^{t_{1} \times t_{2}},~~ l_{1} = 1,2, \hdots, L_{1}
\end{equation}
where $\text{Mat}_{t_{1},t_{2}}\left[d\right]$ is an operator that reshapes a column vector $d \in \mathbb{R}^{t_{1}t_{2}}$ to a matrix $\boldsymbol{W} \in \mathbb{R}^{t_{1} \times t_{2}}$ and $\mathcal{Q}_{l_{1}}\left\{\boldsymbol{\text{XX}}^{\top}\right\}$ denotes the $l_{1}^{\text{th}}$ principal eigenvector of $\boldsymbol{\text{XX}}^{\top}$. The $L_{1}$ principal eigenvectors capture the main variation of the centralized image patches in the training data. Similar to a convolutional neural network we stack multiple stages of the sparse PCA filters to extract higher level features.

\subsubsection{Second Stage (GS-PCA)}
We repeat the same process as in first stage. Let the $l^{\text{th}}$ filter output of first stage be
\begin{equation}
\label{eq:10}
    I_{i}^{1,l_{1}} \dotequal I_{i} \Conv \boldsymbol{W}_{l_{1}}^{1}, \qquad i = 1, 2, \hdots, N
\end{equation}
where $\Conv$ denotes 2D convolution and boundary of the images $I_{i}$ are zero padded before convolution. Similar to the first stage we collect all the overlapping image patches of the convolved image $I_{i}^{1,l_{1}}$, subtract the patch mean from each patch and obtain the centralized matrix $\boldsymbol{\Bar{\text{Y}}}_{i}^{l_{1}} = \left[\boldsymbol{\Bar{\text{y}}}_{i,1}, \boldsymbol{\Bar{\text{y}}}_{i,2}, \hdots, \boldsymbol{\Bar{\text{y}}}_{i,\Tilde{u}\Tilde{v}}\right]$, where $\boldsymbol{\Bar{\text{y}}}_{i,j}$ is the $j^{\text{th}}$ mean subtracted image patch in $I_{i}^{1,l_{1}}$. We define $\boldsymbol{\text{Y}}^{l_{1}} = [\boldsymbol{\Bar{\text{Y}}}_{1}^{l_{1}}, \boldsymbol{\Bar{\text{Y}}}_{2}^{l_{1}}, \hdots \boldsymbol{\Bar{\text{Y}}}_{N}^{l_{1}}]$ as the matrix containing all the mean subtracted patches of the $l^{\text{th}}$ filter output and concatenate $\boldsymbol{\text{Y}}^{l}$ for all filter outputs as
\begin{equation}
\label{eq:11}
    \boldsymbol{\text{Y}} = \left[\boldsymbol{\text{Y}}^{1}, \boldsymbol{\text{Y}}^{1}, \hdots, \boldsymbol{\text{Y}}^{L_{1}}\right] ~\in ~\mathbb{R}^{t_{1}t_{2} \times L_{1}N\Tilde{u}\Tilde{v}}
\end{equation}
Once again we solve (\ref{eq:7}) with $\boldsymbol{\text{Y}}$ as the input. The solution to the minimization problem in (\ref{eq:7}) are the $L_{2}$ principal eigenvectors of $\boldsymbol{\text{YY}}^{\top}$. The sparse PCA filters of the second stage are then obtained as
\begin{equation}
\label{eq:12}
    \boldsymbol{W}_{l_{2}}^{2} \dotequal \text{Mat}_{t_{1},t_{2}}\left[\mathcal{Q}_{l_{2}}\left\{\boldsymbol{\text{YY}}^{\top}\right\}\right] \in \mathbb{R}^{t_{1} \times t_{2}},~~ l_{2} = 1,2, \hdots, L_{2}.
\end{equation}
For each input image $I_{i}^{1,l_{1}}$ of the second stage, there will be $L_{2}$ output images of size $u \times v$ generated as
\begin{equation}
\label{eq:13}
    I_{i}^{2,l_{2}} \dotequal \left\{I_{i}^{1,l_{1}} \Conv \boldsymbol{W}_{l_{2}}^{2}\right\}^{L_{2}}_{l_{2} = 1}
\end{equation}
After the second stage we will obtain $L_{1}L_{2}$ output images. It is easy to repeat the above process to build more (sparse PCA) stages if a deeper architecture is needed.

\subsubsection{Binary Quantization (Hashing)}
For each of the $L_{1}$ input images $I_{i}^{1,l_{1}}$ presented to the second stage we obtain $L_{2}$ real-valued output images $I_{i}^{2,l_{2}}$. We binarize these outputs and obtain $\{H(I_{i}^{1,l_{1}} \Conv \boldsymbol{W}_{l_{2}}^{2})\}^{L_{2}}_{l_{2} = 1}$, where $H\left(\cdot\right)$ is a Heaviside step (like) function, which has a value of 1 for positive entries and zero otherwise. Around each pixel, we view the vector of $L_{2}$ binary bits as a decimal number, thus converting the $L_{2}$ outputs in $I_{i}^{2,l_{2}}$ into a single integer-valued ``image"
\begin{equation}
\label{eq:14}
    T_{i}^{l_{1}} \dotequal \sum\limits_{l_{2} = 1}^{L_{2}} 2^{l_{2} - 1} H\left(I_{i}^{1,l_{1}} \Conv \boldsymbol{W}_{l_{2}}^{2}\right),
\end{equation}
which has pixel values in the range $\left[0, 2^{L_{2}}-1\right]$.

\subsubsection{Block-wise Histograms}
We partition each of the $L_{1}$ ``images" $T_{i}^{l_{1}}, l_{1} = 1, 2, \hdots, L_{1}$ into $G$ distinct blocks, compute the histogram (with $2^{L_{2}}$ bins) of the decimal values in each block and concatenate all $G$ histograms into a single vector denoting it as $G_{\text{hist}}(T_{i}^{l_{1}})$. After such an encoding process the ``feature" of the input image $I_{i}$ is then defined to be the set of block-wise histograms, i.e.,
\begin{equation}
\label{eq:15}
    f_{i} \dotequal \left[G_{\text{hist}}\left(T_{i}^{1}\right), \hdots, G_{\text{hist}}\left(T_{i}^{L_{1}}\right)\right] ~\in~ \mathbb{R}^{\left(2^{L_{2}}\right)L_{1}G}.
\end{equation}
We use overlapping blocks to build the feature vector for each input image $I_{i}$ as it helps in retaining most amount of the information.

We train a linear support vector machine (SVM) classifier \cite{cortes95support} using the feature vector $f_{i}$ obtained for each input image $I_{i}$ from the GS-PCANet in order to classify cancer lesions versus normal tissues on H\&E stained histological lung slides.

\subsection{Classifying Color Images}
\label{sec:mthdc}

There are several options to extend the proposed GS-PCANet method to be able to extract features for classifying color images. In this work, we follow the approach described in \cite{gurcan09histo, chan15pcanet} and apply the proposed GS-PCANet to each of the red, blue, and green channels to obtain multichannel sparse PCA filters, that are then used to extract features for classifying the color images. 

\section{Experiments and Results}
\label{sec:res}

In this section we evaluate our proposed  GS-PCANet image classification algorithm with other open-source histopathology image classification methods: SpPCANet method for image classification \cite{dutta20sparse}, multiple clustered instance learning (MCIL) for histopathology image classification \cite{xu14weakly}, saliency-based dictionary learning (SDL) \cite{sarkar18sdl}, analysis-synthesis learning with shared features (ASLF) \cite{li20anal}, patch-based convolutional neural network (PCNN) \cite{hou16patch}, encoded local projections (ELP) for histopathology image classification \cite{tizhoosh18represent}, and weakly supervised deep learning (WSDL) for whole slide tissue classification \cite{campanella19clinical}. We evaluate these seven methods using commonly used detection/classification measures: precision ($\text{P}$), recall ($\text{R}$), detection accuracy, $F_{\beta}$-score, Tanimoto coefficient ($\text{T}$), and the receiver operating characteristic (ROC) curves along with the area under the curve (AUC). 

The Precision $\text{P}$ and recall $\text{R}$ (a.k.a. true positive rate or sensitivity) are given by
\begin{eqnarray}
\label{eq:16}
    \text{P} = \frac{\text{TP}}{\text{TP} + \text{FP}}, \quad \text{R} = \frac{\text{TP}}{\text{TP} + \text{FN}}
\end{eqnarray}
where TP is the number of true positive classifications, FP is the number of false positive classifications, and FN is the number of false negative classifications. The false positive rate (a.k.a. complement of specificity) is defined as $\text{FP}/(\text{FP} + \text{FN})$. An ROC curve is a plot of the true positive rate versus the false positive rate. The detection accuracy is defined as ($\text{TP} + \text{TN})/(\text{TP} + \text{FP} + \text{TN} + \text{FN})$ .

The $F_{\beta}$-score is defined by
\begin{eqnarray}
\label{eq:17}
    F_{\beta} = \left(1 + \beta^{2}\right)\frac{\text{P\,R}}{\left(\beta^{2}\text{P}\right) + \text{R}}
\end{eqnarray}
We use $F_{1}$ (i.e., $\beta = 1$) as this is the most common choice for this type of evaluation \cite{ram16size}.

Tanimoto coefficient, also known as Tanimoto distance in statistics, is defined as
\begin{eqnarray}
\label{eq:18}
    \text{T} = \frac{\text{TP}}{\text{M} + \text{N} - \text{TP}}
\end{eqnarray}
where $\text{M}$ is the number of detected individual tumors by an automated algorithm and $\text{N}$ is the actual number of individual tumors in the image.

The AUC is the average of precision $\text{P}(\text{R})$ over the interval ($0 \leq \text{R} \leq 1$), where $\text{P}(\text{R})$ is a function of recall $\text{R}$. It is given by
\begin{eqnarray}
\label{eq:19}
    \text{AUC} = \int_{0}^{1} \text{P}(\text{R})\,d\text{R}.
\end{eqnarray}
The best detection algorithm among several alternatives is commonly defined as the one that maximizes the Tanimoto coefficient, AUC, and the $F_{\beta}$-score.

\subsection{Dataset}
\label{sec:resa}

The proposed method was mainly developed with the goal of identifying individual tumors in H\&E stained whole slide histopathology lung images obtained from an inducible \textit{K-ras}$^{\textit{G12D}}$ lung cancer model. The images were produced using a digital slide scanner (Super COOLSCAN 5000 ED Digital Slide Scanner; Nikon Corporation) with a $1\times$ objective lens (level-$0$ pixel size: $0.52\mu\text{m} \times 0.52\mu\text{m}$). In our experiments, the size of each image acquired is approximately $3000 \times 3000$ pixels. Our dataset consists of a total of 67 whole slide histopathology lung images obtained from 32 non-tumor-bearing mice and 35 mice with visible tumors. A careful manual delineation of the borders of the individual tumors within the 35 images was performed by an expert and considered as ground truth for subsequent analysis. We divide each image in our dataset into non-overlapping image patches of size $20 \times 20$ pixels consisting of a total of 52,487 cancer lesion patches and 1,455,023 normal patches. 

\subsection{Experimental Setup}
\label{sec:resb}

We used a total of 15 non-tumor-bearing mice images and 15 images with visible tumors for training the compared algorithms, consisting of a total of 21,934 cancer lesion patches and 653,092 normal patches. Our test dataset consists of 17 non-tumor-bearing mice images and 20 images with visible tumors consisting of a total of 30,553 cancer lesion patches and 801,931 normal patches. The hyper-parameters of the GS-PCANet algorithm include the filter size ($t_{1}, t_{2}$), the number of stages, the number of filters in each stage ($L_{1}, L_{2}$), and the block size for the local histograms in the output stage. The optimal values for these parameters were automatically selected on a validation set (randomly chosen from within the training data), using the ROC curves by varying one parameter at a time while keeping the others fixed and choosing that value of the parameter that maximizes the AUC of the ROC curve. The parameters of the GS-PCANet were set to $t_{1} = t_{2} = 5$, $L_{1} = 9$, $L_{2} = 45$, and, a histogram block size of $8 \times 8$.
\begin{figure*}[!t]
\includegraphics[width= 7.14in, height=3.31in]{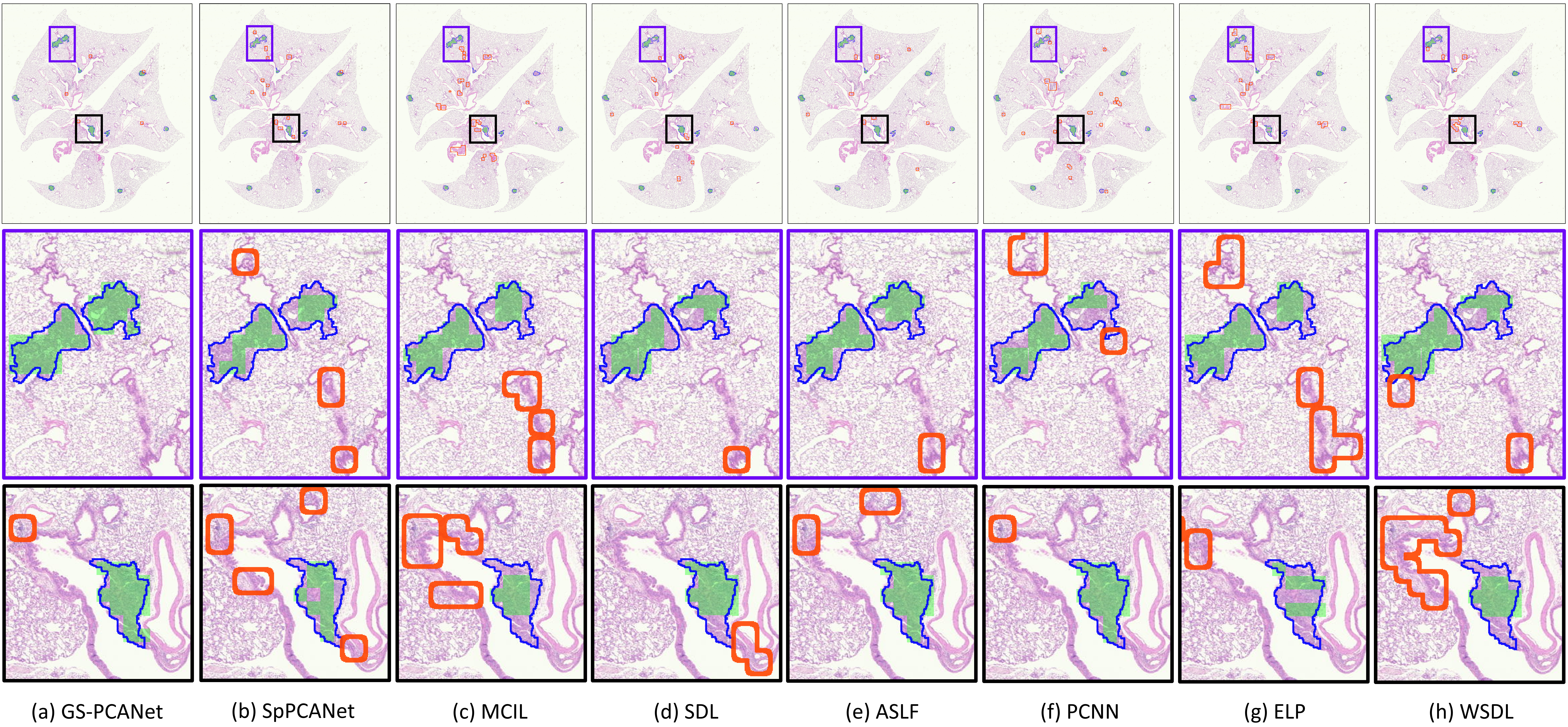}
\caption{Detection results on a representative image containing visible tumors in our test dataset using: (a) GS-PCANet, (b) SpPCANet, (c) MCIL, (d) SDL, (e) ASLF, (f) PCNN, (g) ELP, and (h) WSDL. The true borders delineated by an expert of each individual tumor in the image are shown in blue, the true positives patches identified by each method are shown in green and the false positives of each method are bordered in red in the color version of this paper. False negatives are those regions within the blue-bordered individual tumors that are not shaded in green. Results on the entire image are shown in row 1, and two zoomed regions are shown zoomed in rows 2 and 3.}
\label{fig3}
\vspace{-6mm}
\end{figure*}
\begin{figure}[!t]
\includegraphics[width= 3.7in, height=2.77in]{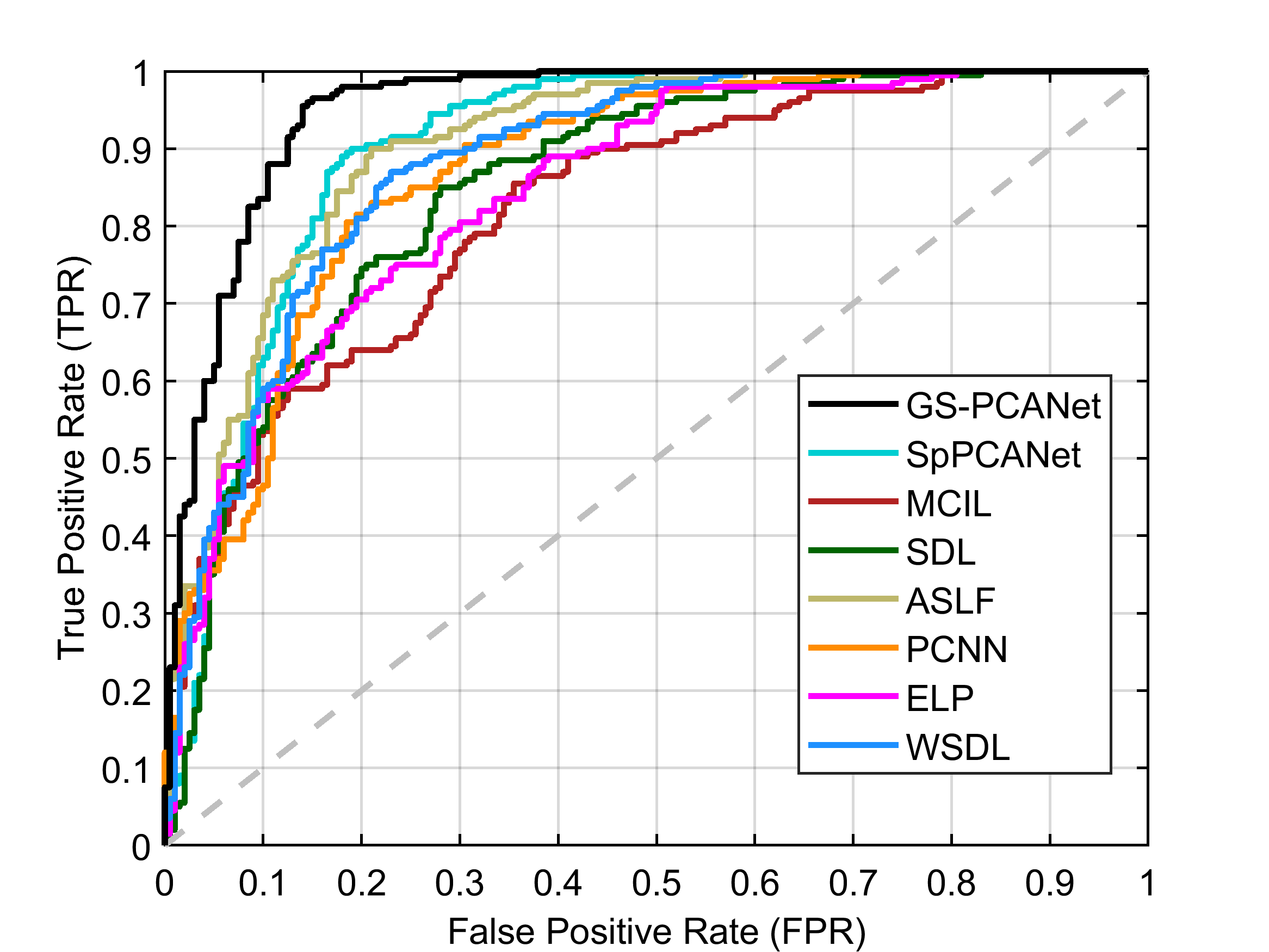}
\caption{ROC curve of image patch classification as cancerous or healthy for different methods.}
\label{fig4}
\vspace{-6mm}
\end{figure}

\subsection{Qualitative Results}
\label{sec:resc}

Fig.~\ref{fig3} shows the qualitative detection results for an example image containing visible tumors from our test dataset. Fig.~\ref{fig3}(a) shows that the proposed GS-PCANet method detects most of tumor regions correctly with very few false positives and false negatives. Fig.~\ref{fig3}(d) shows that the ASLF method is also able to identify the tumor regions well, but detects more false positives than the GS-PCANet method. The SpPCANet, MCIL, and WSDL methods have many misclassifications (with blood vessels being identified as tumors) as shown in Figs.~\ref{fig3}(b), (c) and (g), respectively. The ELP method splits a single tumor into three tumors (see Fig.~\ref{fig3}(g) row 3), with many false positives. The SDL, PCNN, and ELP methods miss large parts of individual tumors, i.e., have many false negatives as shown in Fig.~\ref{fig3}(d), (f), and (g), respectively. Visually it is clear that the proposed GS-PCANet method accurately detects both large and small individual tumors within the whole slide image with very few false positives and false negatives. This is of great significance for those studying oncogenesis, progression, and metastasis because the robustness of the algorithm to the size of the tumor reduces the likelihood that the algorithm will mislabel cases containing only small tumors.
\begin{table*}[!t]
\caption{Mean Performance (and Standard Deviation) for Various Algorithms}
\begin{center}
\vspace{-3mm}
\renewcommand{\arraystretch}{1.4}
\begin{tabular}{ >{\centering} m{2.5cm}| >{\centering} m{2.1cm} >{\centering}m{2.1cm} >{\centering}m{2.1cm} >{\centering}m{2.1cm} >{\centering}m{2.1cm} >{\centering}m{2.1cm}}
\hline 
 \rowcolor[gray] {0.8}\textbf{Method} & \textbf{Precision ($\text{P}$)} & \textbf{Recall ($\text{R}$)} & \textbf{$F_{\beta}$-score} & \textbf{Tanimoto Coefficient ($\text{T}$)} & \textbf{Detection Accuracy} & \textbf{AUC} \tabularnewline \hline
\textbf{GS-PCANet} & \textbf{0.872} ~ (\textbf{0.013}) & \textbf{0.955} ~ (\textbf{0.019}) & \textbf{0.912} ~ (\textbf{0.015}) & \textbf{0.903} ~ (\textbf{0.010}) & \textbf{0.908} ~ (\textbf{0.008})  & \textbf{0.951} $\pm$ \textbf{0.011}\tabularnewline
\rowcolor[gray] {0.95}\textbf{SpPCANet} \cite{dutta20sparse} & 0.841 ~ (0.019) & 0.870 ~ (0.025) & 0.855 ~ (0.022) & 0.836 ~ (0.014) & 0.853 ~ (0.015) & 0.907 $\pm$ 0.017\tabularnewline
\textbf{MCIL} \cite{xu14weakly} & 0.719 ~ (0.022) & 0.780 ~ (0.015) & 0.748 ~ (0.031) & 0.762 ~ (0.019) & 0.738 ~ (0.026) & 0.821 $\pm$ 0.013\tabularnewline
\rowcolor[gray] {0.95}\textbf{SDL} \cite{sarkar18sdl} & 0.752 ~ (0.024) & 0.850 ~ (0.031) & 0.798 ~ (0.025) & 0.801 ~ (0.017) & 0.785 ~ (0.011) & 0.849 $\pm$ 0.021\tabularnewline
\textbf{ASLF} \cite{li20anal} & 0.811 ~ (0.028) & 0.900 ~ (0.019) & 0.853 ~ (0.021) & 0.829 ~ (0.030) & 0.845 ~ (0.018) & 0.903 $\pm$ 0.022\tabularnewline
\rowcolor[gray] {0.95}\textbf{PCNN} \cite{hou16patch} & 0.807 ~ (0.039) & 0.815 ~ (0.031) & 0.811 ~ (0.032) & 0.796 ~ (0.023) & 0.810 ~ (0.024) & 0.871 $\pm$ 0.039\tabularnewline
\textbf{ELP} \cite{tizhoosh18represent} & 0.761 ~ (0.023) & 0.750 ~ (0.018) & 0.756 ~ (0.021) & 0.739 ~ (0.027) & 0.758 ~ (0.023) & 0.844 $\pm$ 0.014\tabularnewline
\rowcolor[gray] {0.95}\textbf{WSDL} \cite{campanella19clinical} & 0.798 ~ (0.030) & 0.785 ~ (0.028) & 0.823 ~ (0.031) & 0.821 ~ (0.035) & 0.818 ~ (0.028) & 0.882 $\pm$ 0.041\tabularnewline \hline
\end{tabular}
\end{center}
\label{table1}
\vspace{-8mm}
\end{table*}

\subsection{Quantitative Results}
\label{sec:resd}

\begin{figure}[!t]
\includegraphics[width= 3.7in, height=2.77in]{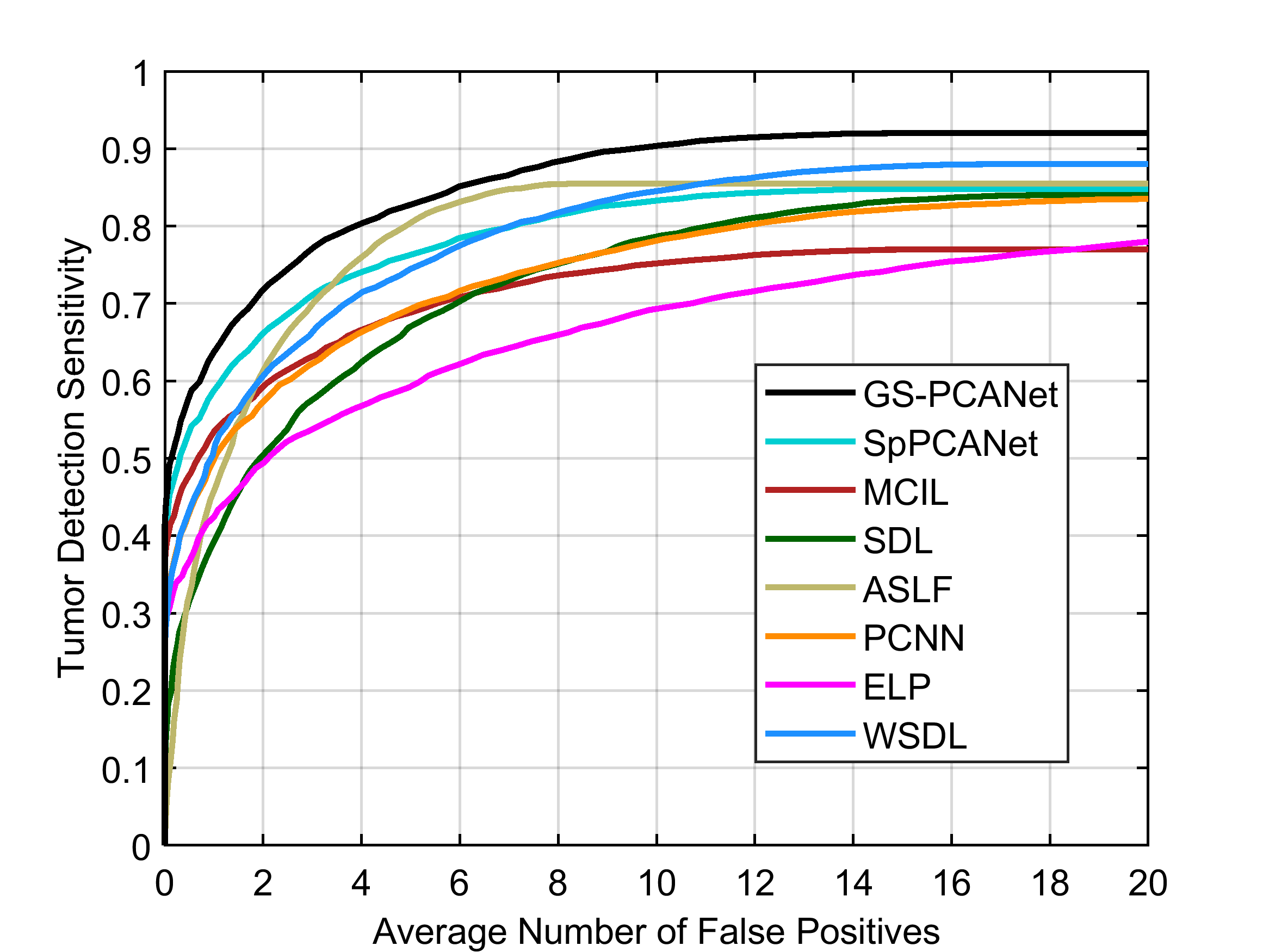}
\caption{FROC curve of different methods for the individual tumor detection task within an entire image.}
\label{fig5}
\vspace{-3mm}
\end{figure}

We compared the quantitative performance of the automated methods at the image patch level and for the task of individual tumor detection within an entire image as well. Fig.~\ref{fig4} shows the ROC curves of all automated methods at the image patch level on the test dataset. From Fig.~\ref{fig4}, we observe that our proposed GS-PCANet method exhibits the most favorable trade-off in terms of accurate detection while maintaining low false positive rate in comparison to the other automated methods. Table~\ref{table1} shows the quantitative performance of the compared methods for the task of individual tumor detection within the histopathology images in the test dataset. Table~\ref{table1} shows that the detection accuracy of the proposed GS-PCANet method is much higher than the other competing algorithms. From Table~\ref{table1}, we also observe that the $F_{\beta}$-score, and Tanimoto coefficient ($\text{T}$) of the proposed method are the highest among the compared algorithms. Table~\ref{table1} also provides the AUC values and their 95\% confidence intervals corresponding to the ROC curves in Fig.~\ref{fig4} for each method. We observe from the AUC values that the GS-PCANet method outperforms the alternatives. In addition to the metrics in Table~\ref{table1}, we also computed the free receiver operating characteristics curves (FROC) \cite{ram16size} for all the compared algorithms. Fig.~\ref{fig5} shows that the proposed GS-PCANet method has better detection accuracy compared to the other automated methods at all points along the FROC curve. This shows that the proposed method detects the individual tumors within these images better than the other compared methods.

The confusion matrix corresponding to competing methods for our test dataset is provided in Table~\ref{table2}. From Table~\ref{table2}, we observe that our proposed GS-PCANet method outperforms competing dictionary learning methods as well as the deep learning methods. This success is attributed to the ability of our proposed GS-PCANet method to capture both the local and the global features associated with both normal and cancerous regions within the images, which the other compared methods do not address.
\begin{figure}[!t]
\includegraphics[width= 3.7in, height=2.77in]{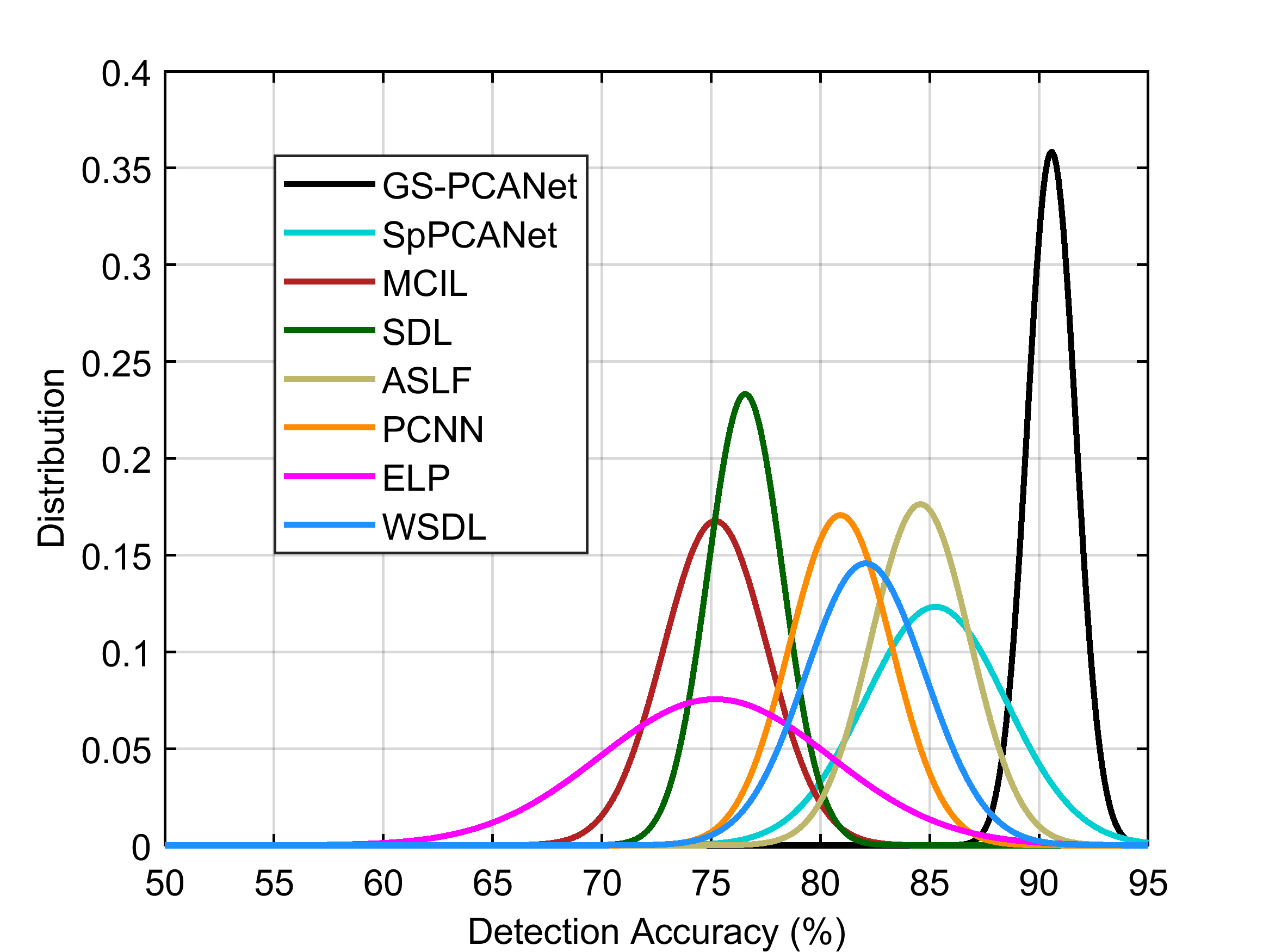}
\caption{Selection bias plot showing the distribution of detection accuracy over ten different training choices of image patches for the compared methods.}
\label{fig6}
\vspace{-3mm}
\end{figure}
\begin{table}[!t]
\caption{Confusion Matrix (\%)}
\begin{center}
\vspace{-3mm}
\renewcommand{\arraystretch}{1.4}
\begin{tabular}{ >{\centering} m{1.4cm}| >{\centering} m{1.5cm}| >{\centering}m{1.5cm}| >{\centering}m{2.5cm}}
\hline 
 \cellcolor[gray] {0.8}\textbf{Class} & \cellcolor{blue!25}\textbf{Cancerous} & \cellcolor{red!25}\textbf{Healthy} & \cellcolor[gray] {0.8}\textbf{Method}\tabularnewline \hline
 \cellcolor{blue!25} & \textbf{87.21} & \textbf{12.79} & \cellcolor[gray] {0.95}\textbf{GS-PCANet}\tabularnewline
 \cellcolor{blue!25} & 84.06 & 15.94 & \cellcolor[gray] {0.95}\textbf{SpPCANet} \cite{dutta20sparse}\tabularnewline
 \cellcolor{blue!25} & 71.89 & 28.11 & \cellcolor[gray] {0.95}\textbf{MCIL} \cite{xu14weakly}\tabularnewline
 \cellcolor{blue!25} & 75.22 & 24.78 & \cellcolor[gray] {0.95}\textbf{SDL} \cite{sarkar18sdl}\tabularnewline
 \cellcolor{blue!25}\textbf{Cancerous} & 81.08 & 18.92 & \cellcolor[gray] {0.95}\textbf{ASLF} \cite{li20anal}\tabularnewline
 \cellcolor{blue!25} & 80.69 & 19.31 & \cellcolor[gray] {0.95}\textbf{PCNN} \cite{hou16patch}\tabularnewline
 \cellcolor{blue!25} & 76.14 & 23.86 & \cellcolor[gray] {0.95}\textbf{ELP} \cite{tizhoosh18represent}\tabularnewline
 \cellcolor{blue!25} & 79.81 & 20.19 & \cellcolor[gray] {0.95}\textbf{WSDL} \cite{campanella19clinical}\tabularnewline \hline
 \cellcolor{red!25} & \textbf{\textcolor{white}{0}4.97} & \textbf{95.03} & \cellcolor[gray] {0.95}\textbf{GS-PCANet}\tabularnewline
 \cellcolor{red!25} & 13.47 & 86.53 & \cellcolor[gray] {0.95}\textbf{SpPCANet} \cite{dutta20sparse}\tabularnewline
 \cellcolor{red!25} & 24.04 & 75.96 & \cellcolor[gray] {0.95}\textbf{MCIL} \cite{xu14weakly}\tabularnewline
 \cellcolor{red!25} & 17.24 & 82.76 & \cellcolor[gray] {0.95}\textbf{SDL} \cite{sarkar18sdl}\tabularnewline
 \cellcolor{red!25}\textbf{Healthy} & 11.24 & 88.76 & \cellcolor[gray] {0.95}\textbf{ASLF} \cite{li20anal}\tabularnewline
 \cellcolor{red!25} & 18.69 & 81.31 & \cellcolor[gray] {0.95}\textbf{PCNN} \cite{hou16patch}\tabularnewline
 \cellcolor{red!25} & 24.63 & 73.37 & \cellcolor[gray] {0.95}\textbf{ELP} \cite{tizhoosh18represent}\tabularnewline
 \cellcolor{red!25} & 16.04 & 83.96 & \cellcolor[gray] {0.95}\textbf{WSDL} \cite{campanella19clinical}\tabularnewline \hline
 \end{tabular}
\end{center}
\label{table2}
\vspace{-5mm}
\end{table}

\subsection{Statistical Analysis}
\label{sec:rese}

To investigate the robustness of training or selection bias for each automated method, we obtain the detection performance for 10 different choices of training image patches (the number of training images were fixed), using the rest of the image patches as test image patches. The detection accuracy for each training run was fit to a Gaussian probability density function (pdf) and plotted in Fig.~\ref{fig6}. From Fig.~\ref{fig6}, we observe that the mean our proposed GS-PCANet curve is much higher than the competing methods indicating superior average detection accuracy. Even more crucial is the spread/variance of our GS-PCANet curve is smaller than its alternatives indicating highly desirable robustness to the particular choice of training image patches.
\begin{figure}[!t]
\includegraphics[width= 3.7in, height=2.77in]{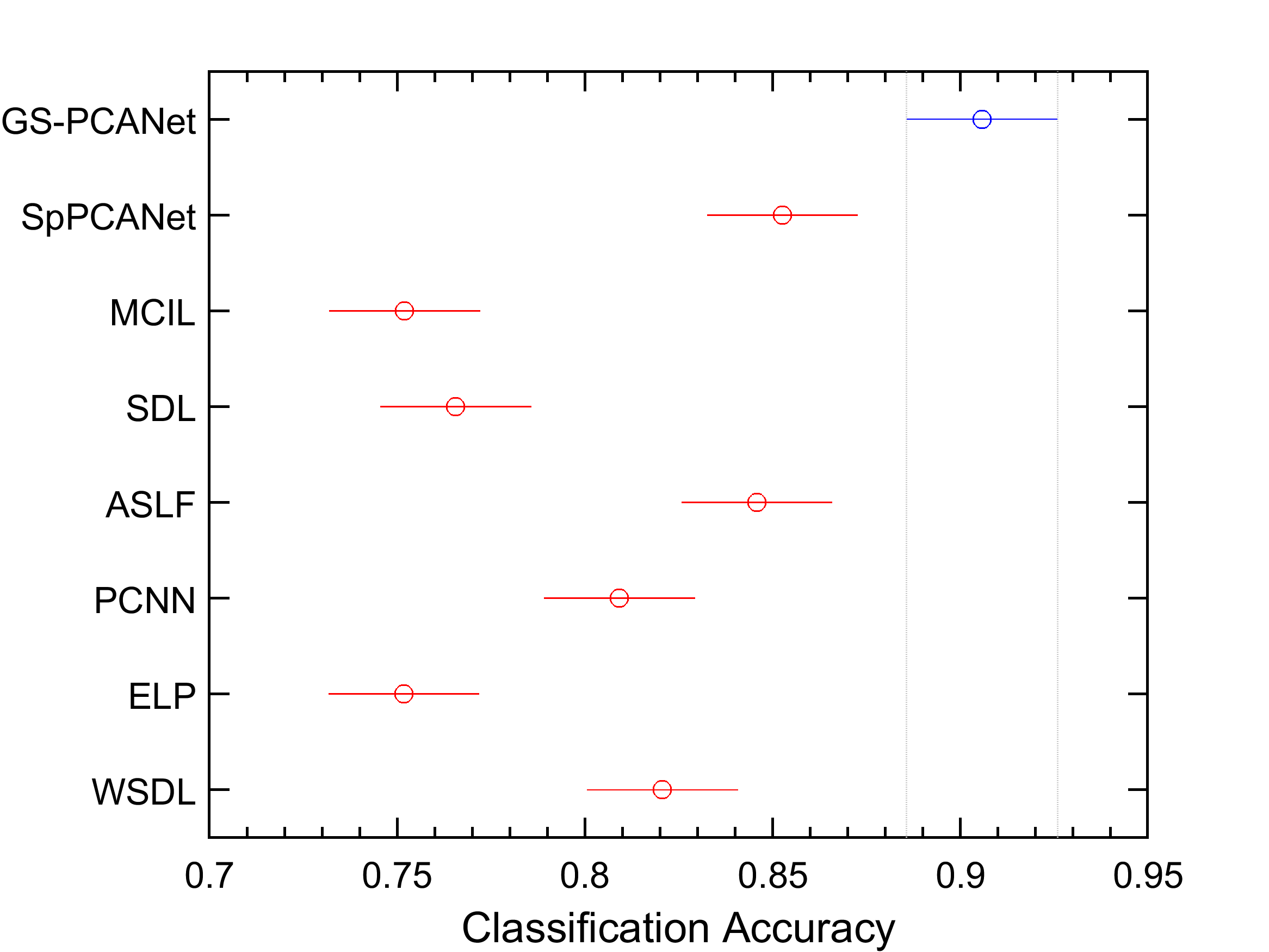}
\caption{Comparison of the proposed GS-PCANet method and other state-of-the-art alternatives by a two-way ANOVA. Values reported by ANOVA (using MATLAB function \emph{anova2}) across the methods are $S\!S = 0.2103, df = 7, M\!S = 0.0300, F = 36.27, p \ll 1\text{e-5}$, indicating that the improved accuracy of the proposed GS-PCANet method is statistically significant. The intervals shown represent 95\% confidence intervals of the detection accuracies for the proposed method (blue) and the competing methods (red).}
\label{fig7}
\vspace{-4mm}
\end{figure}

We also performed a balanced two-way analysis of variance (ANOVA) \cite{hogg87engine} on the detection accuracies in the selection-bias experiment for all the methods. Fig.~\ref{fig7} shows these comparisons using a post-hoc Tukey range test \cite{hogg87engine}. Fig.~\ref{fig7} shows that the performance of the GS-PCANet method is significantly separated from its competing alternatives. $p$-values of the proposed GS-PCANet method compared with other state-of-the-art methods are observed to be much less than $1\times10^{-5}$, emphasizing the fact that the GS-PCANet method is more effective.

\subsection{Computational Complexity}
\label{res:secf}

Here we show computational complexity of the GS-PCANet method by considering a two stage network. For each stage in the GS-PCANet, forming the mean subtracted image patch matrix $\boldsymbol{\text{X}}$ has a computational complexity of $\mathcal{O}\left(t_{1}t_{2}\Tilde{u}\Tilde{v}\right)$; the inner product $\boldsymbol{\text{XX}}^{\top}$ in (\ref{eq:9}) has a complexity of $\mathcal{O}\left(\left\{t_{1}t_{2}\right\}^{2}\Tilde{u}\Tilde{v}\right)$; the computational complexity of the eigen decomposition with graph-regularization is $\mathcal{O}\left(\left\{t_{1}t_{2}\right\}^{3}\right)$. The sparse PCA filter convolution has a complexity of $\mathcal{O}\left(L_{i}t_{1}t_{2}uv\right)$ at stage $i$. The block-wise histogram computation has a complexity of $\mathcal{O}\left(uvGL_{2}\right)$. With $\Tilde{u} = u - \left(t_{1} - 1\right)$, $\Tilde{v} = v - \left(t_{2} - 1\right)$, and assuming $uv \gg \max\left(t_{1}, t_{2}, L_{1}, L_{2}, G\right)$, the overall complexity of GS-PCANet is
\begin{equation}
\label{eq:20}
    \mathcal{O}\left(uvt_{1}t_{2}\left\{L_{1} + L_{2}\right\} + uv\left\{t_{1}t_{2}\right\}^{2}\right).
\end{equation}
The computational complexity in (\ref{eq:20}) applies to both the training and testing phase of GS-PCANet because the extra computation burden during training is the eigen decomposition, which can be ignored when $uv \gg \max\left(t_{1}, t_{2}, L_{1}, L_{2}, G\right)$.  
\begin{table}[!t]
\caption{Mean Run Time (and Standard Deviation)}
\label{table3}
  \begin{center}
  \vspace{-4mm}
	\renewcommand{\arraystretch}{1.4}
	\begin{tabular}{>{\centering} m{2.7cm}  >{\centering} m{2.15cm}  >{\centering} m{2.15cm} }
	\hline
	\rowcolor[gray] {0.8}\textbf{Method} & \textbf{Training Time (HH:MM:SS)} & \textbf{Run Time (Std. Dev.) in Sec.}  \tabularnewline \hline
	 \textbf{GS-PCANet} & 00:21:09 & 11.14 (3.09) \tabularnewline
	 \textbf{SpPCANet} \cite{dutta20sparse} & 00:20:53 & 15.21 (1.41) \tabularnewline
	 \textbf{MCIL} \cite{xu14weakly}    & 18:25:06 & 66.35 (14.36) \tabularnewline
	 \textbf{SDL} \cite{sarkar18sdl} & 01:22:41 & 46.11 (4.51) \tabularnewline	
	 \textbf{ASLF} \cite{li20anal} & 01:49:27 & 19.39 (5.15) \tabularnewline
	 \textbf{PCNN} \cite{hou16patch}   & 19:27:55 & 39.47 (15.22) \tabularnewline
	 \textbf{ELP} \cite{tizhoosh18represent}   & 04:38:03 & 71.44 (9.40) \tabularnewline
	 \textbf{WSDL} \cite{campanella19clinical}   & 21:44:17 & 10.31 (6.02) \tabularnewline\hline
	\end{tabular}
    \vspace{-6mm}
   \end{center}
\end{table}
\begin{figure}[!b]
\vspace{-6mm}
\includegraphics[width= 3.7in, height=2.77in]{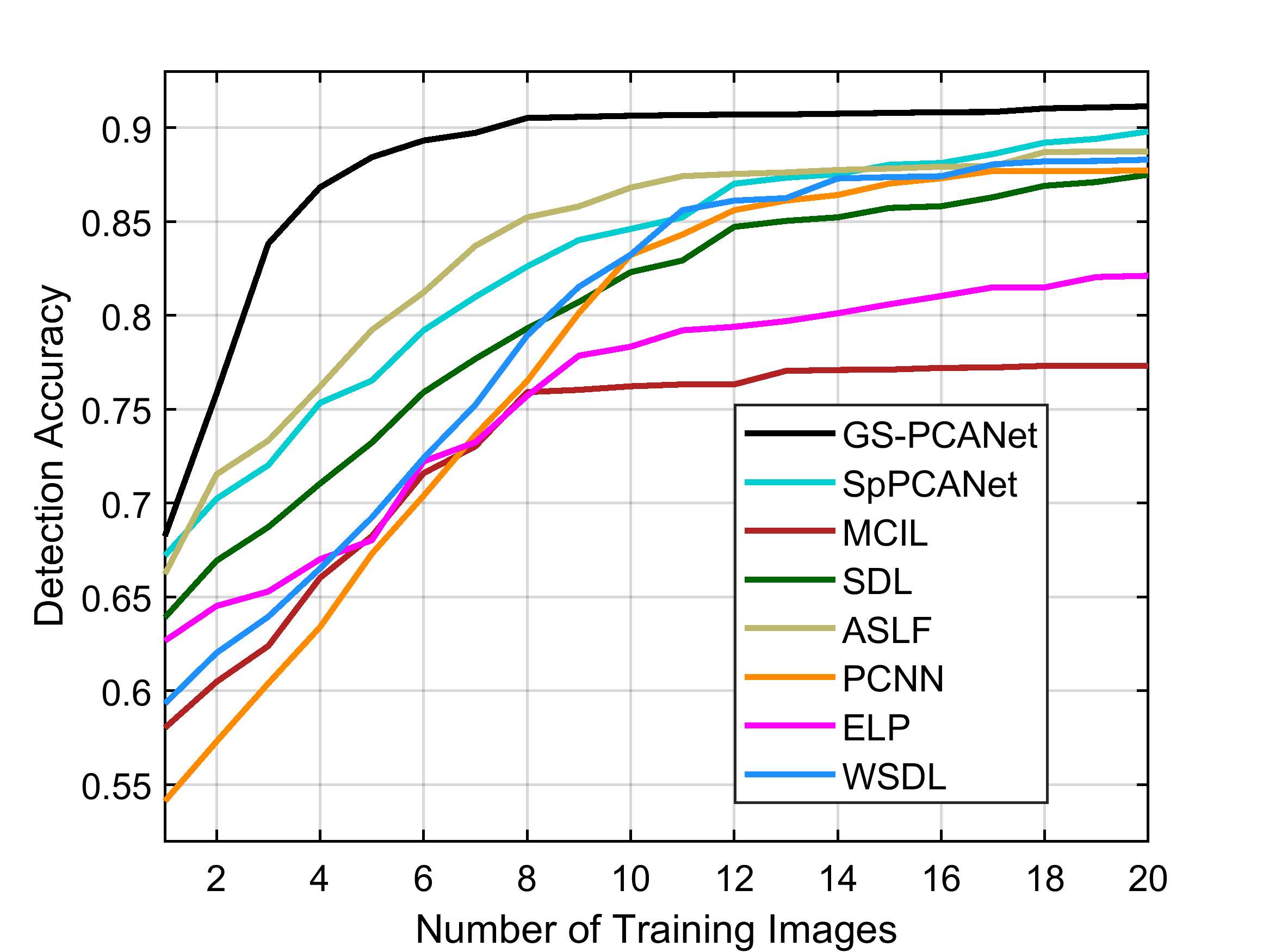}
\caption{Detection accuracy as a function of the number of training images for the competing methods.}
\label{fig8}
\vspace{-3mm}
\end{figure}

We compared the mean inference run time, namely, the time required to classify all the image patches in a single test image for each of the competing algorithms. Table~\ref{table3} shows the mean and standard deviation of the run time each method takes to classify an entire image. From Table~\ref{table3}, we observe that the proposed GS-PCANet method runs 0.83 seconds slower than the WSDL method, but is on average faster than all the other methods. The SDL and ASLF methods classify the test image patch by reconstructing them from the learned dictionaries and thus take more time to execute at test time. The ELP algorithm finds the Radon transformation of each test image patch at various orientations, thereby taking more time to classify each test image patch. The MCIL method integrates the clustering of multiple subtypes of a single class into the MIL classification framework, thus requiring more run time compared to the other methods. In Table~\ref{table3} we also report the training time required to train each of the competing algorithms. From Table~\ref{table3}, we observe that the proposed GS-PCANet method and the SpPCANet method take roughly about 21 minutes to train, where as the other methods take about 3 to 62 times more time to train a good model. The small training time of the GS-PCANet method is attributed to the low computational complexity of the method.

\subsection{Impact on Number of Training Images}
\label{res:secg}

In this section, we show the practicality and applicability of the proposed GS-PCANet method in medical imaging tasks where we have very few data to learn from. Whereas in all other experiments we trained on 15 images each, from both classes, in this experiment we varied the number of training images (from 1 to 20) for all the competing methods and computed detection accuracy of these methods. Fig.~\ref{fig8} shows the detection accuracy of all the competing algorithms on the test dataset of 27 images (12 non-tumor images and 15 images with visible tumors). From Fig.~\ref{fig8}, we observe that the proposed GS-PCANet method trained with as few as 8 images achieves a high detection accuracy of 91\%, whereas the other methods are able to achieve a maximum detection accuracy of only about 89\% and also require as much as 20 training images. This shows that the proposed GS-PCANet method can produce a good model for image classification with less training data. 

\section{Discussion and Conclusion}
\label{dis-conl}

Tumor burden in histopathological sections is difficult to assess by manual evaluation, as well as by prior automated tumor detection algorithms. To solve this problem, our proposed machine learning algorithm uses a cascaded graph-based sparse PCA transform followed by PCA binary hashing and block-wise histograms to obtain features within image patches. These features are then used to classify an image patch as cancerous or healthy using a linear SVM classifier. Our approach differs from earlier learning-based methods based on deep learning \cite{hou16patch,campanella19clinical}, instance learning \cite{xu14weakly,tizhoosh18represent} or dictionary learning \cite{sarkar18sdl,li20anal} for histopathology image classification. Like many deep learning methods, the network parameters, such as the number of stages, the filter size, and the number of filters, need to be optimized and fixed for our GS-PCANet method. Once these parameters are fixed, training the GS-PCANet is extremely simple and efficient because the filter learning in GS-PCANet does not require regularized parameters or require numerical optimization solvers. Moreover, the GS-PCANet consists of only linear operations at each stage with a non-linearity applied only at the output stage, which makes the method more interpretable than other deep learning methodologies.

The GS-PCANet method was first validated with respect to detection accuracy using ROC curves and the AUC of the ROC curve. Second, the algorithm was validated with respect to detection accuracy using the precision, recall, $F_{\beta}$-score, Tanimoto coefficient, FROC curves, and the confusion matrix. Tables~\ref{table1}\,\&\,\ref{table2} show that the proposed GS-PCANet method performs the best among the compared methods for histopathology image classification. Fig.~\ref{fig3} shows that the proposed GS-PCANet method qualitatively performs the best in comparison to the other methods. Further, Fig.~\ref{fig6} shows that the GS-PCANet method has superior average detection accuracy and is more robust to the choice of training images compared to the other methods. We also show the low computational complexity of the GS-PCANet method and compare the training and inference run times for all the methods. Table~\ref{table3} shows that the GS-PCANet method is relatively very fast to learn a good model in comparison to other methods. Finally, Fig.~\ref{fig8} shows that the proposed method requires less data to learn a good model.  
\begin{figure}[!t]
\includegraphics[width= 3.5in, height=2.62in]{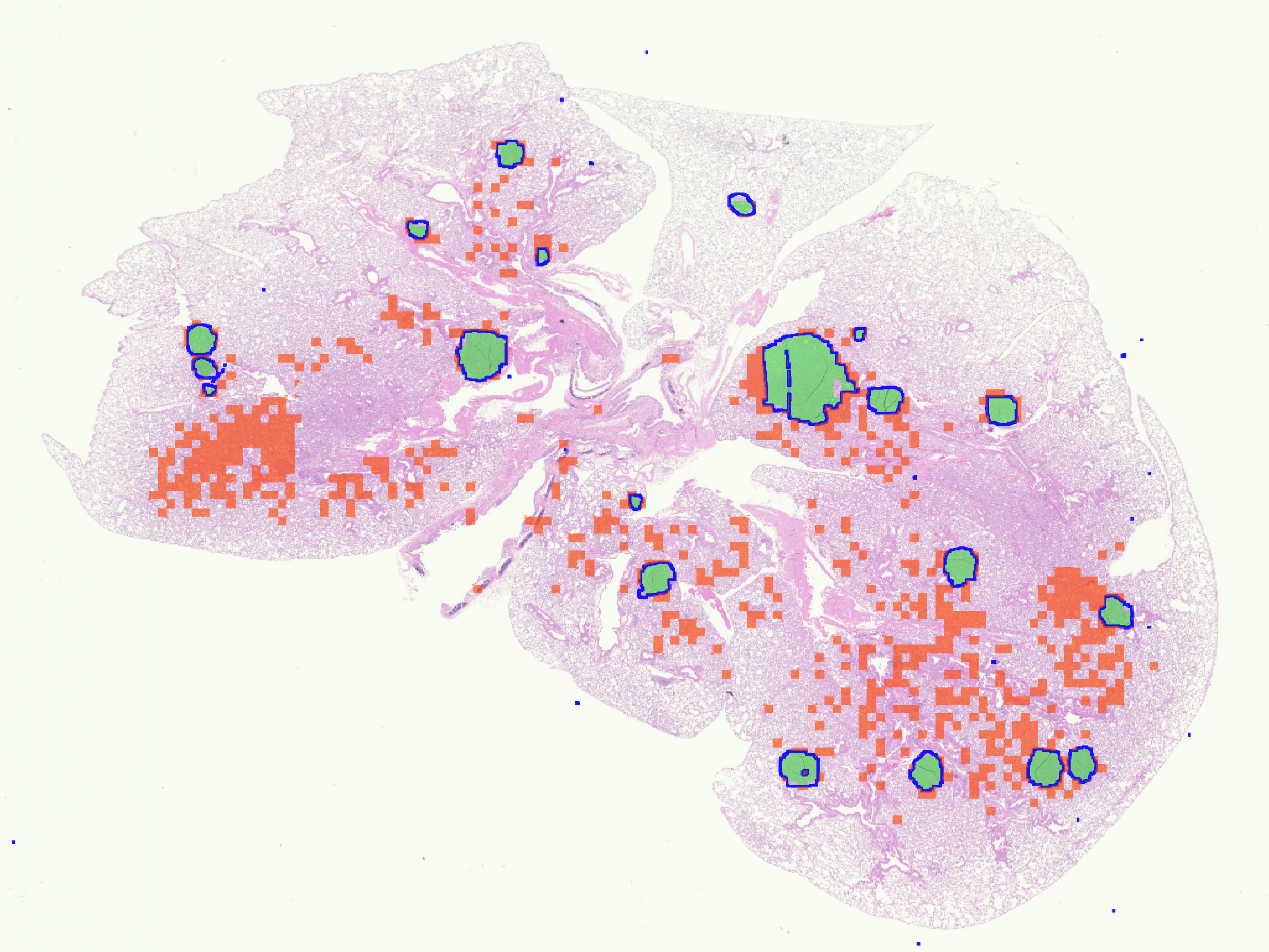}
\caption{Example of detection errors produced by our algorithm on an image with visible tumors. The true borders delineated by an expert of each individual tumor in the image are shown in blue, the true positive and the false positive image patches are shown in green and red, respectively, in the color version of this paper (the image is better viewed in zoomed mode).}
\label{fig9}
\vspace{-4mm}
\end{figure}

Next, we present some inherent limitations of the automated methods for tumor detection. Fig.~\ref{fig9} shows an example case of an image containing individual tumors where all algorithms including our algorithm fail to produce optimum detection results. In Fig.~\ref{fig9} we observe that even though the algorithm has detected all the individual tumors, i.e., the true positive image patches shown in green color, it has also detected many false positive image patches shown in red color. On close examination, we see that the false positive image patches within the image look very similar to cancerous image patches. This could be due to the fact that there is not enough resolution in this image to differentiate between the cancerous and healthy image patches, or this histopathology section was captured when some of the underlying cells were transitioning from healthy to being cancerous. 

The proposed detection algorithm uses all the image patches in the training data for obtaining the local structures within the data when computing the graph-based term in (\ref{eq:6}) and (\ref{eq:7}). This adds to the time complexity and results in noise and outlier image patches still being included. However, the algorithm can be modified by linearly clustering the image patches into subgroups and taking these cluster centers to compute the graph term in (\ref{eq:7}). Making this change could further reduce detection errors and also accelerate the algorithm, making it more accurate and efficient at the same time.    

\bibliographystyle{IEEEtran}
\bibliography{IEEEabrv,myref}

%
%
%
%
%
%
%




\end{document}